\documentclass[12pt,a4paper]{article}

\usepackage{epsfig}

\begin{document}


\noindent
DESY 03-175, SFB/CPP-03-48, CPT-2003-P-4604\hfill{\tt hep-lat/0311002}\\
November 2003

\begin{center}
{\LARGE\bf Staggered versus overlap fermions:\\[2mm]
a study in the Schwinger model with $N_f=0,1,2$}
\end{center}

\begin{center}
{\large\bf Stephan D\"urr$\,{}^{a}$}
\hspace{8pt}{\large and}\hspace{8pt}
{\large\bf Christian Hoelbling$\,{}^{b}$}
\\[10pt]
${}^a\,$DESY Zeuthen, Platanenallee 6, D-15738 Zeuthen, Germany\\
${}^b\,$Centre de Physique Th\'eorique, Case 907, CNRS Luminy,
F-13288 Marseille Cedex 9, France\\
\end{center}

\begin{abstract}
\noindent
We study the scalar condensate and the topological susceptibility for a
continuous range of quark masses in the Schwinger model with $N_f=0,1,2$
dynamical flavors, using both the overlap and the staggered discretization.
At finite lattice spacing the differences between the two formulations become
rather dramatic near the chiral limit, but they get severely reduced, at the
coupling considered, after a few smearing steps.
\end{abstract}


\newcommand{\pad}{\partial}
\newcommand{\pas}{\partial\!\!\!/}
\newcommand{\Dsl}{D\!\!\!\!/\,}
\newcommand{\Psl}{P\!\!\!\!/\;\!}
\newcommand{\hqu}{\hbar}
\newcommand{\ovr}{\over}
\newcommand{\til}{\tilde}
\newcommand{\pri}{^\prime}
\renewcommand{\dag}{^\dagger}
\newcommand{\<}{\langle}
\renewcommand{\>}{\rangle}
\newcommand{\gaf}{\gamma_5}
\newcommand{\lap}{\triangle}
\newcommand{\trc}{\rm tr}

\newcommand{\al}{\alpha}
\newcommand{\be}{\beta}
\newcommand{\ga}{\gamma}
\newcommand{\de}{\delta}
\newcommand{\ep}{\epsilon}
\newcommand{\ve}{\varepsilon}
\newcommand{\ze}{\zeta}
\newcommand{\et}{\eta}
\renewcommand{\th}{\theta}
\newcommand{\vt}{\vartheta}
\newcommand{\io}{\iota}
\newcommand{\ka}{\kappa}
\newcommand{\la}{\lambda}
\newcommand{\rh}{\rho}
\newcommand{\vr}{\varrho}
\newcommand{\si}{\sigma}
\newcommand{\ta}{\tau}
\newcommand{\ph}{\phi}
\newcommand{\vp}{\varphi}
\newcommand{\ch}{\chi}
\newcommand{\ps}{\psi}
\newcommand{\om}{\omega}

\newcommand{\psb}{\overline{\psi}}
\newcommand{\etb}{\overline{\eta}}
\newcommand{\psd}{\psi^{\dagger}}
\newcommand{\etd}{\eta^{\dagger}}
\newcommand{\beq}{\begin{equation}}

\newcommand{\eeq}{\end{equation}}
\newcommand{\bdm}{\begin{displaymath}}
\newcommand{\edm}{\end{displaymath}}
\newcommand{\bea}{\begin{eqnarray}}
\newcommand{\eea}{\end{eqnarray}}

\newcommand{\mr}{\mathrm}
\newcommand{\mb}{\mathbf}
\newcommand{\Nf}{{N_{\!f}}}
\newcommand{\Nc}{{N_{\!c}}}
\newcommand{\ri}{\mr{i}}
\newcommand{\DW}{D_\mr{W}}
\newcommand{\DN}{D_\mr{N}}


\hyphenation{topo-lo-gi-cal simu-la-tion theo-re-ti-cal mini-mum}


\enlargethispage{10pt}
\vspace*{-6pt}

\section{Introduction}

Recently, the staggered action when coupled to ``HYPed'' backgrounds
\cite{Hasenfratz:2001hp} has attracted renewed interest.
This is mainly due to the cheapness of this formulation which bears the promise
that realistic unquenched simulations can be performed with currently available
resources \cite{Davies:2003ik}.
Moreover, as a remnant of the full $SU(\Nf\!=\!d)_A$ group, the continuous
``two-hop global'' symmetry
\beq
\ch(x)\to\exp\Big(\ri\,\th_A(-1)^{\sum n_\nu}\Big)\ch(x)
\quad,\qquad
\bar\ch(x)\to\bar\ch(x)\exp\Big(-\ri\,\th_A(-1)^{\sum n_\nu}\Big)
\;,
\label{susskind}
\eeq
with $x\!=\!a(n_1,\ldots,n_d)$ protects the fermion mass against additive
renormalization.
On the other hand, taking the square and quartic root of the determinant (to
obtain $\Nf\!=\!2\!+\!1$ dynamical flavors) might spoil the locality of the
action \cite{Jansen}, and the often believed insensitivity to topology could
undermine attempts to push towards the chiral limit.

The other extreme in terms of computational cost is represented by the closely
related domain-wall \cite{domainwall} and overlap \cite{overlap} fermions.
For these actions, the Ginsparg-Wilson relation \cite{Ginsparg:1981bj}
\beq
D\hat{\gaf}+\gaf D=0,\qquad
\hat{\gaf}=\gaf(1-\frac{a}{\rho}D)
\label{ginspargwilson}
\eeq
holds in the massless limit (with $\rho$ for the moment an arbitrary papameter
that will be specified in (\ref{diracoverlap})), implying invariance under the
full chiral symmetry group \cite{Luscher:1998pq} 
\beq
\de\ps=\hat{\gaf}\ps,\qquad\de\psb=\psb\gaf
\label{luscher}
\eeq
at finite lattice spacing, which again excludes additive mass renormalization
and prevents operators in different chiral multiplets from mixing.

Given this situation, we decided to investigate the difference between
staggered and overlap fermions at finite lattice spacing in a simple theory
where the concept of chiral symmetry proves relevant and some interesting
quantities are known analytically.
These criteria are fulfilled by the generalized Schwinger model (QED in 2D with
$\Nf$ massive degenerate fermions), a super-renormalizable theory
\cite{Schwinger:tp} where the scale is set through the dimensionful fundamental
coupling
\beq
g={1\ovr a\sqrt{\beta}}
\label{scale}
\;.
\eeq

We produced 10,000 independent gauge configurations on a
$N\!\times\!N\!=\!20^2$ lattice with the standard (compact) Wilson gauge action
at $\beta\!=\!4$, giving a plaquette of $0.86279(10)$.
For each configuration we determine the complete eigenvalue spectrum of the
massless overlap and staggered operators.
This allows us to compute -- for any given mass -- the condensate and the
fermion determinant, which we use to reweight our observables to $\Nf\!=\!1,2$
\cite{reweighting}.

We define the massive overlap operator as
\beq
D^\mr{ov}_m=(1\!-{am\ovr 2\rho})D^\mr{ov}+m\;,\qquad
aD^\mr{ov}=1+\gaf\,\mr{sign}(a\gaf D^\mr{W}_{-\rho})
\label{diracoverlap}
\eeq
with $D^\mr{W}_{-\rho}$ the Wilson operator at negative mass $-\rho/a$.
We further set $\rho\!=\!1$, which we checked, following
Ref.\,\cite{Hernandez:1998et}, is an almost optimal choice with respect to
locality at our coupling.

Previous work on the Schwinger model using a direct approach for the
computation of the scalar condensate is collected in
\cite{previousstag,previousover,Chandrasekharan:1998em} for staggered,
domain-wall/overlap or both actions, respectively.
We checked that we reproduce the staggered condensate from
\cite{Chandrasekharan:1998em} and the overlap condensate of FHLW/GHR in
\cite{previousover}.


\section{Scalar condensate}

In the chiral limit, the $\Nf=1$ scalar condensate is given by
\cite{Schwinger:tp}
\beq
{\ch_\mr{sca}(m\!=\!0)\ovr g}={e^\ga\ovr2\pi^{3/2}}=0.1599\ldots
\quad[\Nf\!=\!1]\;.
\label{schwinger}
\eeq
For $\Nf=2$ a non-zero value would signal spontaneous symmetry breaking and
therefore violate the Mermin-Wagner-Hohenberg-Coleman theorem
\cite{MerminWagnerHohenbergColeman}.
The prediction how explicit symmetry breaking modifies this zero is
\cite{Smilga:1996pi}
\beq
{\ch_\mr{sca}\ovr g}=0.388\ldots\,\Big({m\ovr g}\Big)^{1/3}
\qquad[\Nf\!=\!2]\;.
\label{smilga}
\eeq

In the staggered formulation we follow \cite{previousstag} and implement the
(1-flavor) condensate through
\beq
{\ch_\mr{sca}\ovr g}=-{1\ovr2}{1\ovr N^2g}\<\bar\ch\ch\>
\;,
\label{condstag}
\eeq
where the purpose of the factor $1/2$ is to compensate the two-fold degeneracy
of the staggered formulation in 2D.
Denoting the eigenvalues of the massless staggered Dirac operator by $\la$
(they show up in complex conjugate pairs with zero real part), the reweighted
condensate is
\beq
{\ch_\mr{sca}\ovr g}={1\ovr2}{1\ovr L^2g}
{\<\det(D^\mr{st}_m)^{\Nf/2}\sum{1\ovr(\la+m)}\>\ovr
\<\det(D^\mr{st}_m)^{\Nf/2}\>}\;,\qquad
\det(D^\mr{st}_m)=\prod(\la\!+\!m)
\;.
\label{condstageig}
\eeq
with $L\!=\!Na$ and the sum and product running over the entire spectrum.
In \cite{previousstag} one finds also the associated free field limit
\beq
{\ch_\mr{sca}\ovr g}={2\ovr N^2}{m\ovr g}\sum_{i,j=1}^{N/2}
{1\ovr(am)^2+\sin(2\pi i/N)^2+\sin(2\pi j/N)^2}
\label{condstagfree}
\eeq
which is, of course, independent of $\Nf$.

For overlap fermions, the scalar condensate is unambiguously defined as
\cite{previousover}
\beq
{\ch_\mr{sca}\ovr g}=-{1\ovr N^2 g}\<\psb\hat{\ps}\>
\;,\qquad
\hat{\ps}={1+\gaf\hat{\gaf}\ovr 2}\ps
\;.
\label{condover}
\eeq
Denoting the eigenvalues of the massless overlap Dirac operator by $\la$, and
remembering that we work with $\rho\!=\!1$, the reweighted condensate is
\beq
{\ch_\mr{sca}\ovr g}={1\ovr L^2g}
{\<\det(D^\mr{ov}_m)^\Nf\sum{1-a\la/2\ovr(1-am/2)\la+m}\>\ovr
\<\det(D^\mr{ov}_m)^\Nf\>}\;,\qquad
\det(D^\mr{ov}_m)=\prod((1\!-\!{am\ovr 2})\la\!+\!m)
\label{condovereig}
\eeq
where the sum runs over the full spectrum.
These eigenvalues occur either in complex conjugate pairs or as isolated
chiral (doubler) modes at $a\lambda\!=\!0\,(2)$.
Finally, one can rewrite (\ref{condovereig}) as
\beq
{\ch_\mr{sca}\ovr g}={1\ovr L^2g}
{\<\det(D^\mr{ov}_m)^\Nf\sum^{\prime}{1\ovr\hat{\la}+m}\>\ovr
\<\det(D^\mr{ov}_m)^\Nf\>}\;,\qquad
\hat{\la}=\left(\la^{-1}-{a\ovr 2}\right)^{-1}
\label{condovereig2}
\eeq
where $a\hat{\la}$ is purely imaginary and the sum excludes the doubler modes at
$a\lambda\!=\!2$.

Due to the (remnant) chiral symmetry of (staggered) overlap fermions, no
subtraction of the condensate is required.
Furthermore, the Schwinger model is super-renormalizable, i.e.\ there is no
renormalization needed at infinite cut-off.
In fact, due to the dimensionful coupling (\ref{scale}), all lattice
renormalization factors have the form $Z=1+O(a^2g^2)$ and therefore are $1$, up
to $O(a^2)$ corrections.
Hence, both the staggered and the overlap discretization yield (for any $\Nf$
and $m\!>\!0$) a finite condensate which is subject to $O(a^2)$ cut-off
effects, without a need for additive or multiplicative renormalization.

A modification which is motivated by what is done in full QCD simulations, is
to consider the Dirac operator on a ``copy'' of an element of the Markov chain
on which one applies one or more APE/HYP smearing steps (which in 2D is the
same).
One then thinks of the operator as one in the original links, which is less
local.
For the staggered action, such a modification preserves the universality class
while considerably reducing the ``taste'' violation \cite{Blum:1996uf}.
Being unaware of any detailed understanding of what the optimum smearing
parameter is, we decided to combine the staple and the original link with equal
weight, which means in a $U(1)$ theory that one takes the arithmetic mean of
the phases.

\bigskip

\begin{figure}
\begin{center}
\epsfig{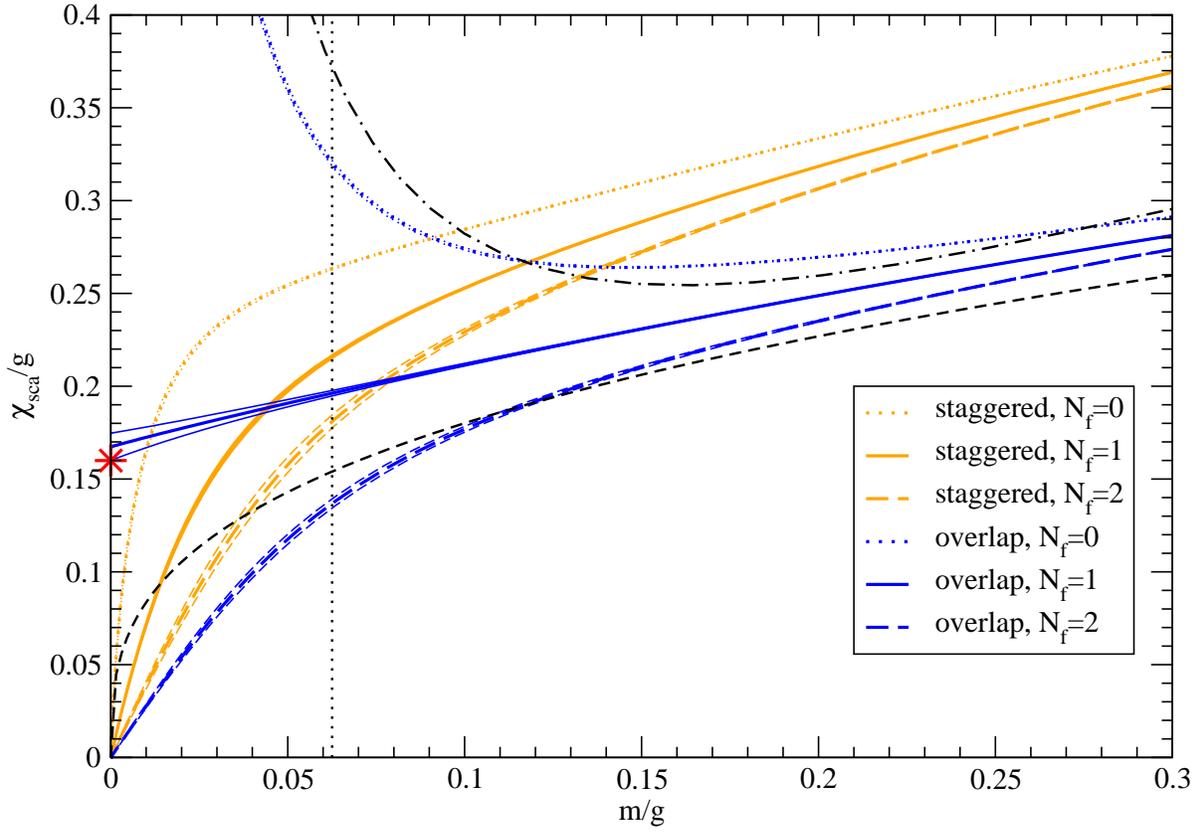}
\epsfig{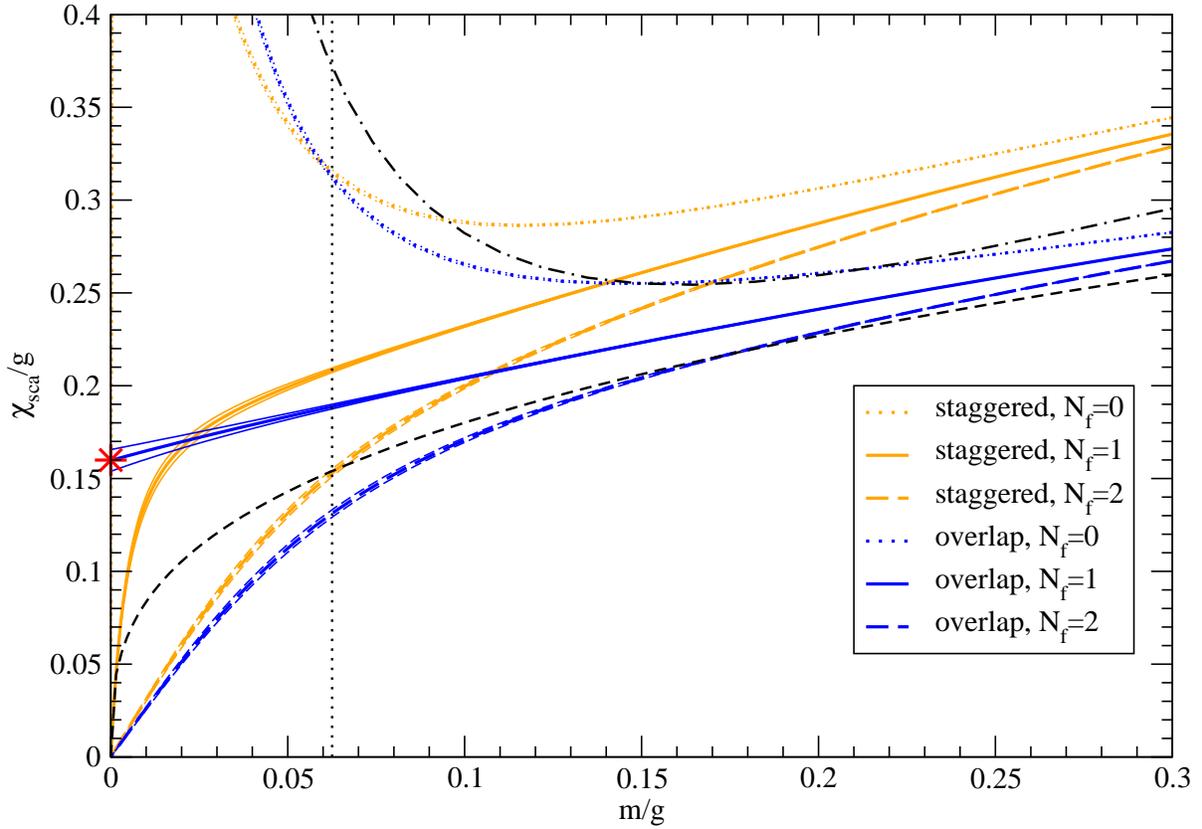}
\end{center}
\vspace{-6mm}
\caption{Scalar condensate after 0 (top) and 1 (bottom) steps of APE/HYP
smearing.}
\label{fig:cond}
\end{figure}

\begin{figure}
\begin{center}
\epsfig{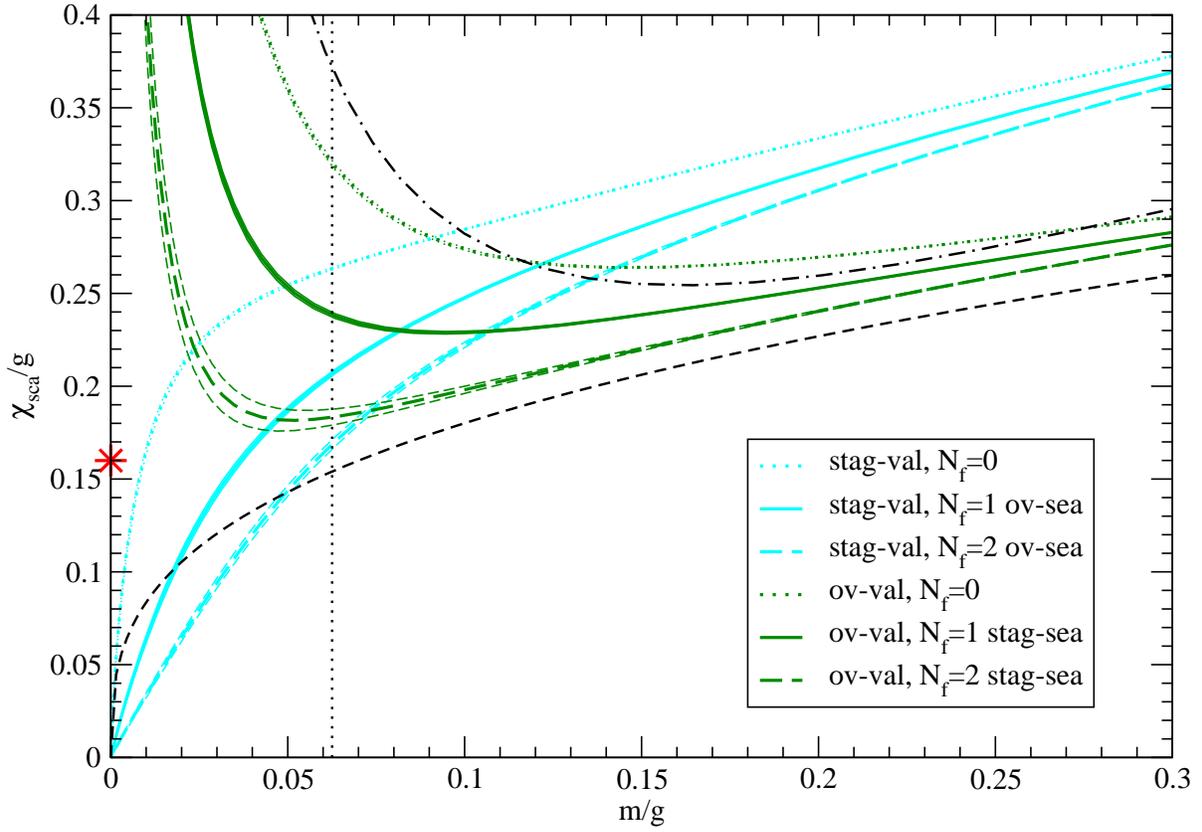}
\epsfig{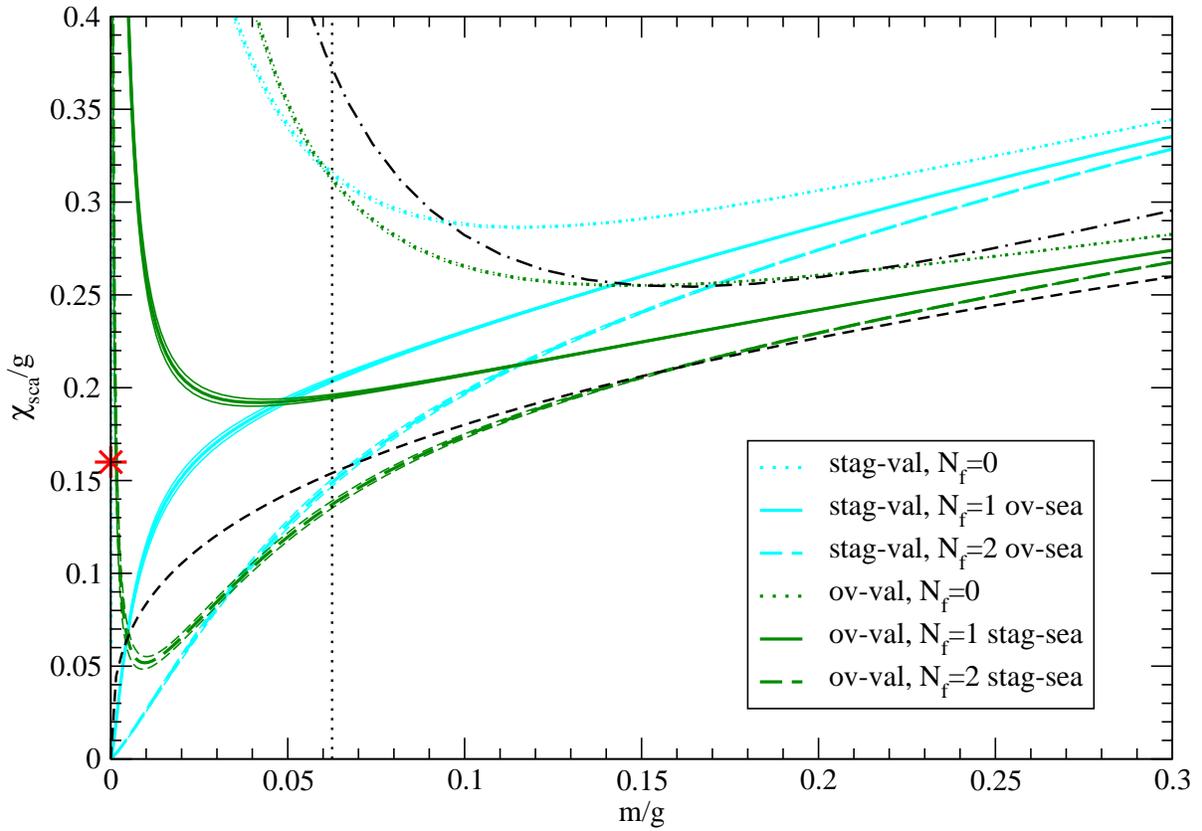}
\end{center}
\vspace{-6mm}
\caption{Hybrid condensate after 0 (top) and 1 (bottom) steps of APE/HYP
smearing.}
\label{fig:hybr}
\end{figure}

\begin{figure}
\begin{center}
\epsfig{file=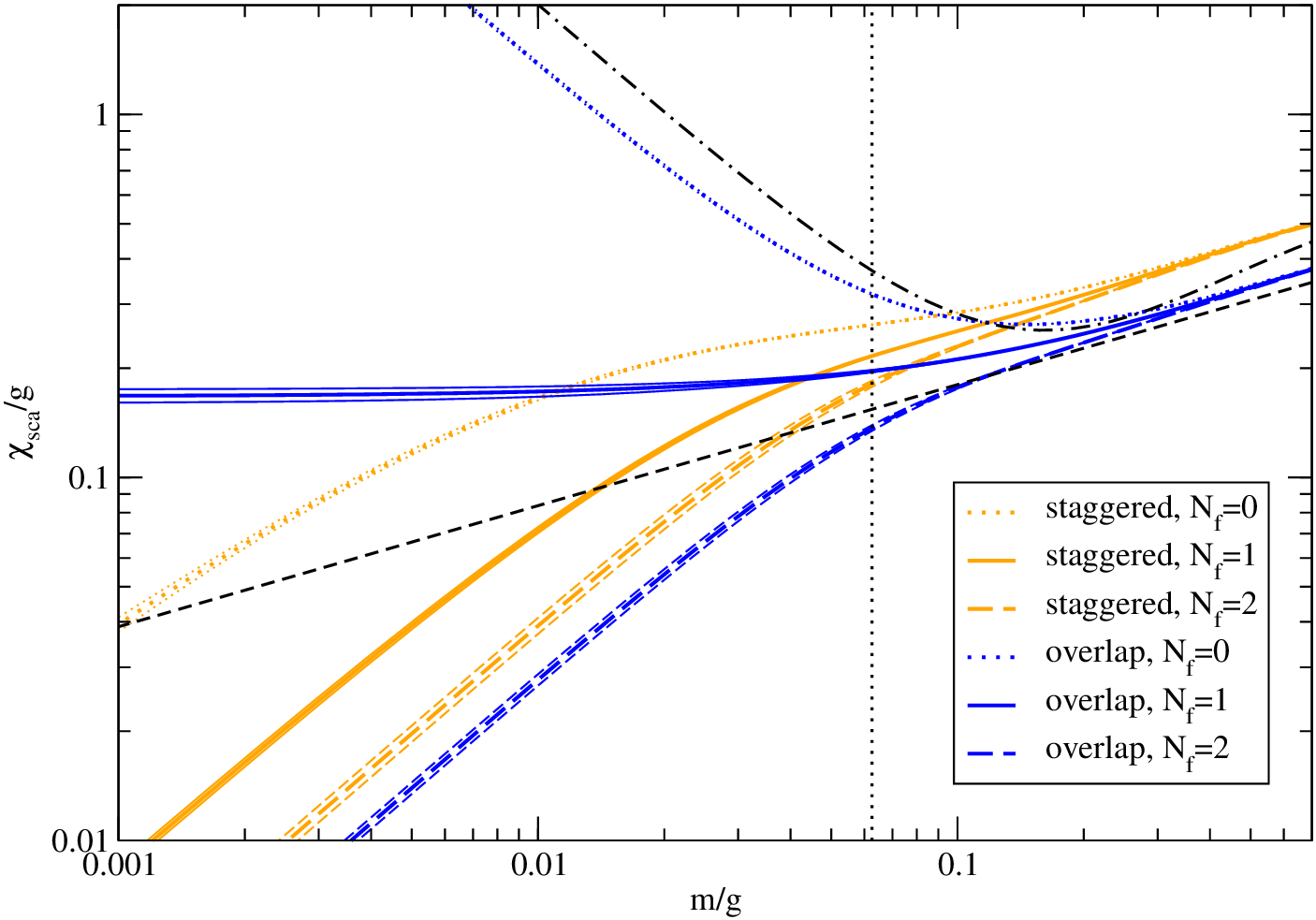,height=11cm}
\epsfig{file=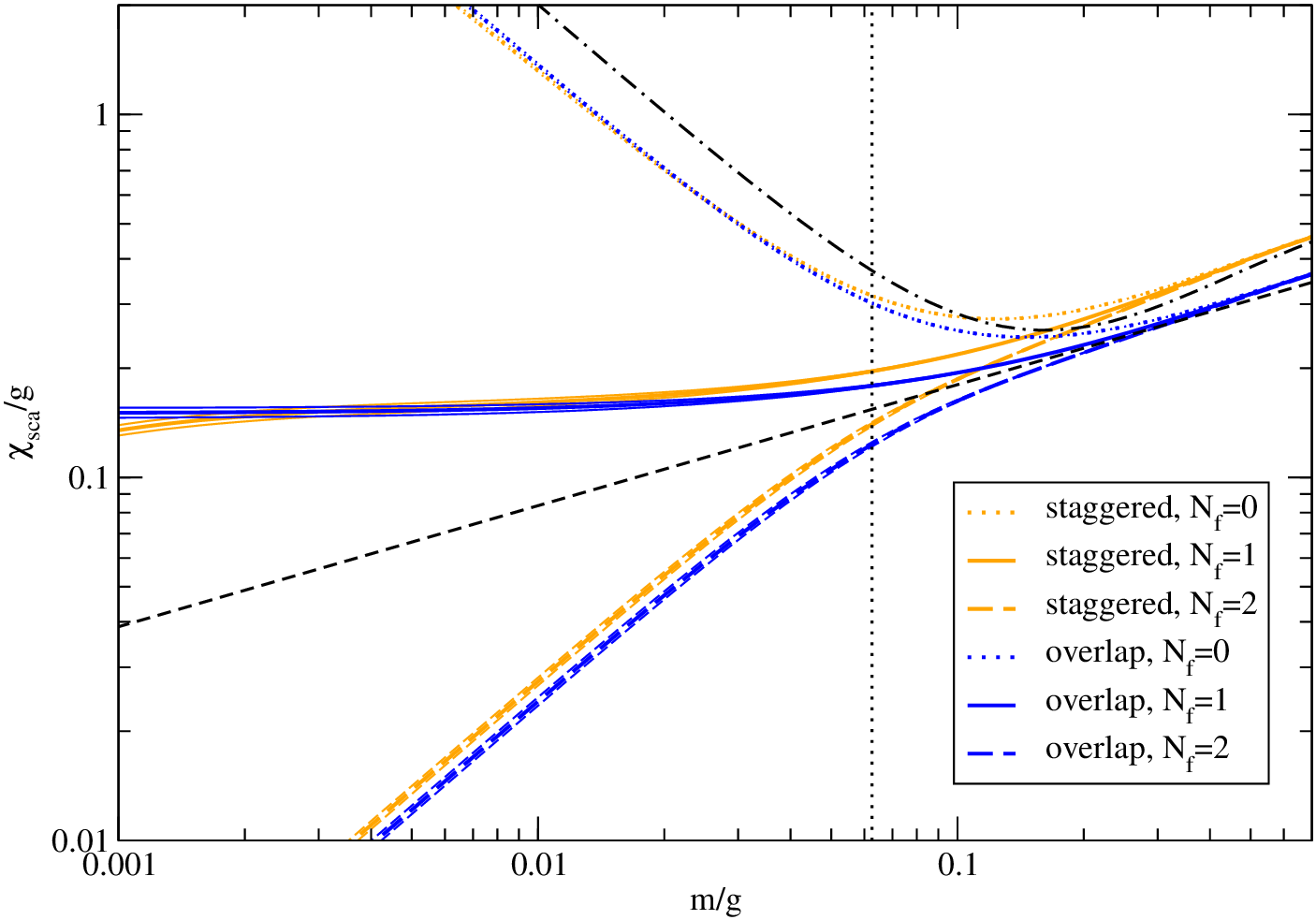,height=11cm}
\end{center}
\vspace{-6mm}
\caption{Scalar condensate after 0 (top) and 3 (bottom) steps of APE/HYP
smearing.}
\label{fig:conl}
\end{figure}

Fig.\,\ref{fig:cond} displays our results for the scalar condensate.
As expected, the overlap condensate exhibits the qualitatively correct behavior
in the chiral limit.
For $\Nf\!=\!0$ it shows the quenched divergence, while for $\Nf\!=\!2$ it
tends to zero indicating the absence of spontaneous chiral symmetry breaking.
In the $\Nf\!=\!1$ case it tends to a constant which seems compatible with the
analytic result (\ref{schwinger}) in infinite volume, marked by an asterisk.
Here, we dare to compare to the infinite volume value, since the results in
\cite{Sachs:en} suggest that for $\Nf\!=\!1$ in a box with
$L\!>\!1/(\sqrt{\pi}g)$ finite volume corrections are exponentially small.
On the other hand, the staggered condensate exhibits a \emph{qualitatively
wrong} behavior in the chiral limit, vanishing for all $\Nf$.
At large masses it differs substantially from the overlap condensate,
presumably due to cutoff effects.
In the intermediate region, there is a smooth turnover with no sign of a
quenched divergence.

The dashed curve is Smilga's infinite volume result (\ref{smilga}), while the
dash-dotted graph represents the free-field expression (\ref{condstagfree}).
Finally, since changing $m$ at fixed box-volume $V$ means that one moves from
the large to the small Leutwyler-Smilga regime, the point where the
Leutwyler-Smilga parameter \cite{Leutwyler:1992yt} 
\beq
x=V\Sigma m
\label{lsp}
\eeq
equals 1 is indicated with a vertical dotted line ($\Sigma$ denotes the
analytical result (\ref{schwinger})).

The real surprise comes when the operators are evaluated after one step of
APE/HYP smearing.
The overlap condensate stays (for any $\Nf$) virtually unchanged, except that
the size of the $O(a^2)$ artefacts is modified.
Formally, the same statement holds true with the staggered action, but from a
quantitative point of view the change is rather dramatic.
In the $\Nf\!=\!1$ case one can ``trust'' it down to much smaller quark masses,
until it finally collapses.
And for $\Nf\!=\!0$ it shows a nice blow-up at moderately small quark masses.
If one really considers the limit $m\!\to\!0$ it still tends to zero, but the
point where this happens is deep within the $\epsilon$ regime ($x\!\ll\!1$)
\cite{Leutwyler:1992yt}.
The effect of additional smearing steps will be discussed below.

\bigskip

Fig.\,\ref{fig:hybr} shows our results for the ``hybrid'' condensate, labeled
after the formulation used for the valence quark.
The ``staggered'' condensate (with $\Nf\!=\!1,2$ sea quarks built from the
overlap determinant) looks virtually unchanged w.r.t.\ Fig.\,\ref{fig:cond}
(the curve with $\Nf\!=\!0$ is identical).
On the other hand, the ``overlap'' condensate (with $\Nf\!=\!1,2$ sea quarks
constructed from the staggered determinant) looks much worse than the original
(true) overlap version; there is a divergence near $m\!=\!0$ for any $\Nf$.
This shows that the failure of the unsmeared (``naive/thin-link'') staggered
formulation cannot be attributed to either the determinant or the valence
prescription alone -- it's the cancellation of the zero in the determinant
and the ``one over zero'' in the propagator which is needed to get the finite
value (\ref{schwinger}) in the 1-flavor case.

Upon applying just one step of smearing the situation improves dramatically;
both ``hybrids'' look qualitatively right down to much smaller quark masses --
but still, eventually the non-faithful representation of the zero-mode(s) in
the staggered part reflects itself in a fake blow-up or drop-down close to
the chiral limit.

\bigskip

Fig.\,\ref{fig:conl} contains our results for the condensate in a log-log
representation.
The overlap data show that the analytic structure in a finite volume is indeed
\beq
\ch_\mr{scal}\propto\left\{
\begin{array}{ll}
1/m&(\Nf\!=\!0)\\
\mr{const}&(\Nf\!=\!1)\\
m&(\Nf\!=\!2)
\end{array}
\right.
\label{nfsplit}
\eeq
near the chiral limit and consistent with
\beq
\ch_\mr{scal}\propto m^{1/3}
\eeq
at large masses, albeit the coefficient in (\ref{smilga}) is
(presumably due to cut-off effects) not reproduced.

The bottom part demonstrates that 3 APE/HYP-smearing steps manage to completely
change the overall picture of the staggered condensate while the overlap
condensate gets rescaled by a factor which differs from 1 only marginally.
Now, the two formulations are in perfect agreement at (moderately) small quark
masses, showing only a mild discrepancy at larger $m/g$: the staggered
condensate seems to tend towards the free-field limit (\ref{condstagfree}),
while the overlap version moves closer to Smilga's prediction (\ref{smilga}).
Eventually, the staggered condensate vanishes in the chiral limit -- even for
$\Nf\!=\!0$. The turnover point, however, is off the scale of the plot.


\section{Selection theorem}

A remarkable feature of the Schwinger model with $\Nf\!=\!1$ is the ``selection
theorem''.
This theorem states that in the chiral limit the non-zero value
(\ref{schwinger}) is formed \emph{exclusively} from the zero-modes of chirality
$\pm1$, which live on backgrounds where the topological charge
\beq
q(A)={g\ovr4\pi}\int\ep_{\mu\nu}F_{\mu\nu}\;dx\;\in{\bf Z}
\label{defq}
\eeq
takes the value $\mp1$.
This statement alludes to a combination of the Atiyah-Singer index theorem and
the vanishing theorem.
The former relates (\ref{defq}) to the index of the Dirac operator, defined
as the difference of the number of negative and positive chirality zero-modes
\beq
\mr{ind}(A)=n_--n_+
\;,
\label{defindex}
\eeq
in a simple manner
\beq
q(A)=\mr{ind}(A)
\;,
\label{indextheorem}
\eeq
while the latter says that in 2D there is no configuration which supports
both positive and negative chirality zero-modes \cite{vanishing}
\beq
\begin{array}{ccc}
n_-\neq0&\Longrightarrow&n_+=0\\
n_+\neq0&\Longrightarrow&n_-=0
\end{array}
\;.
\label{vanishingtheorem}
\eeq
To prove the ``selection theorem'' one starts from the partition function of
the massless theory in the finite volume with periodic boundary conditions
($\Nf\!\geq\!1$) \cite{Sachs:en}
\beq
Z[\bar\et,\et]=N\,\sum_{q\in{\bf Z}}\,\int\!D\!A^{(q)}\;
e^{-{1\ovr4}\!\int\!F^2}\;
\left(\prod_{k=1}^{|q|} (\etb\ps_k)(\psb_k\et)\right)^\Nf
\det{}'(D)^\Nf\;e^{+\!\int\etb S'\et}
\label{partition}
\eeq
from which the condensate is computed by taking \emph{one} derivative w.r.t.\
$\etb$ and $\et$ and then setting the external sources to zero.
The outer sum is over all topological sectors and the product is over the
$|q|$ positive \emph{or} negative chirality zero-modes.
The primes indicate that the determinant and the fermion Green's function
are computed on the subspace orthogonal to the zero-modes.
A non-zero value is obtained \emph{only} if both derivatives hit the prefactor,
leaving nothing but the explicit zero-mode and its conjugate behind.
This can happen only for $\Nf\!=\!1$ and results in a condensate which is
generated \emph{exclusively} by the zero-modes with chirality $\pm1$.
Note that the argument will hold true in QCD, too, once the current body of
numerical evidence in favor of a vanishing theorem in 4D (see Fig.\,1 in
\cite{Blum:2001qg}) has been replaced by a mathematical proof.

As a technical point we mention that on the lattice we define the topological
charge of a background $U$ as the index of the massless overlap operator
\cite{Hasenfratz:1998ri,Luscher:1998pq}
\beq
q(U)=\mr{ind}(U)={1\ovr2}\trc(a\gaf D^\mr{ov})
\label{defind}
\eeq
and use it for both the staggered and the overlap evaluation of the sectoral
condensate.

\bigskip

\begin{figure}
\begin{center}
\epsfig{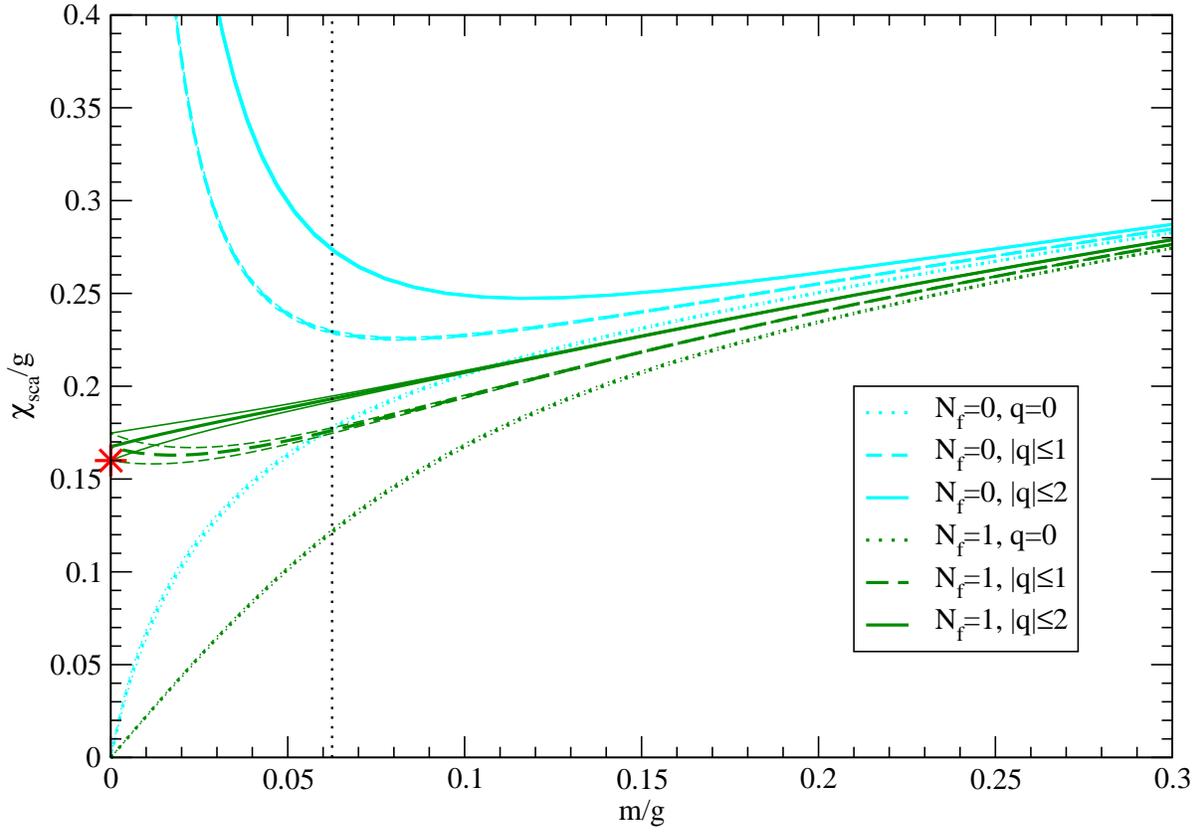}
\epsfig{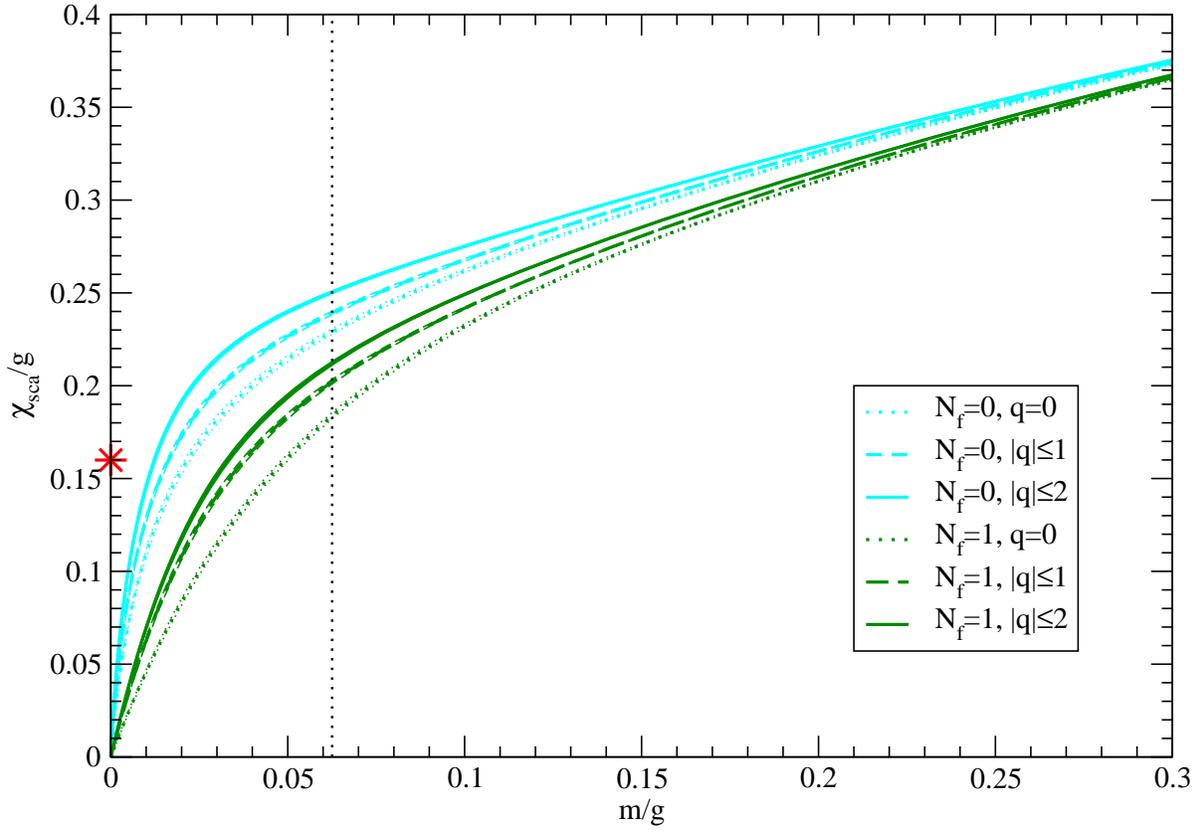}
\end{center}
\vspace{-6mm}
\caption{Sectoral condensate for overlap (top) and staggered quarks without
smearing.}
\label{fig:seco}
\end{figure}

\begin{figure}
\begin{center}
\epsfig{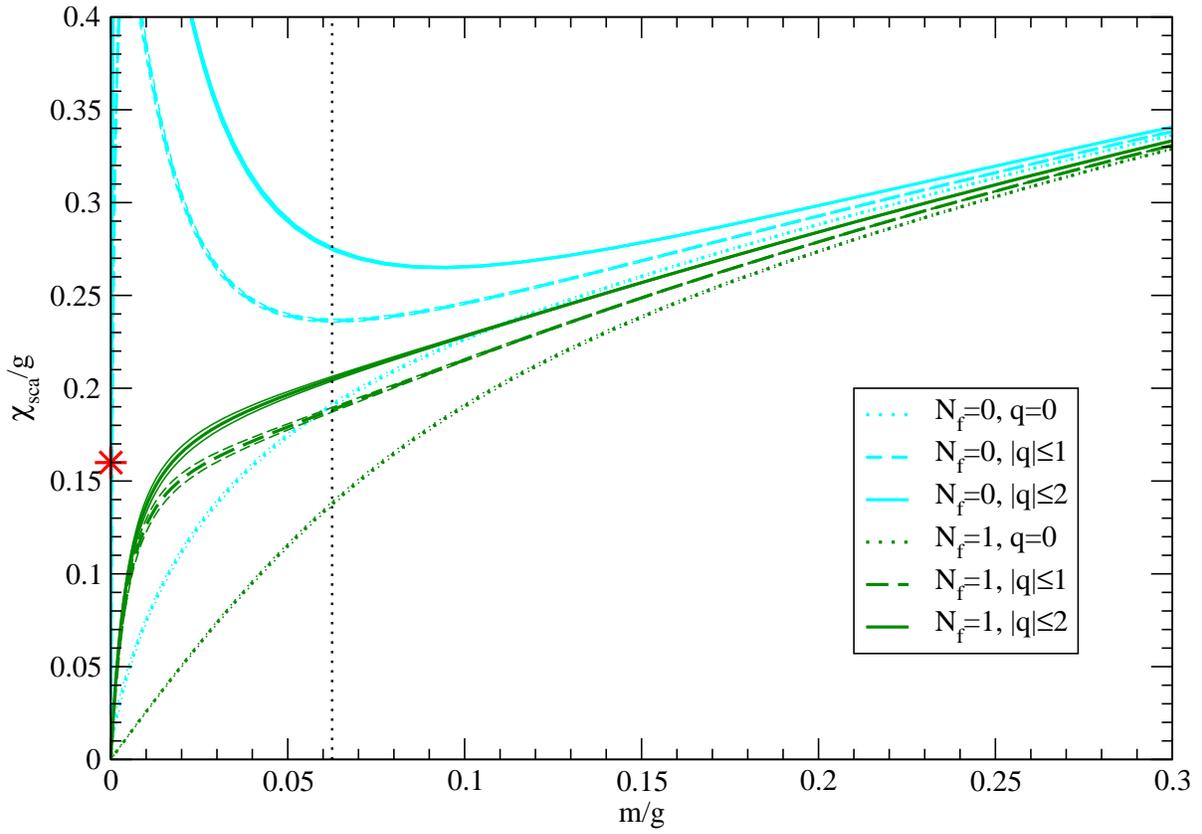}
\epsfig{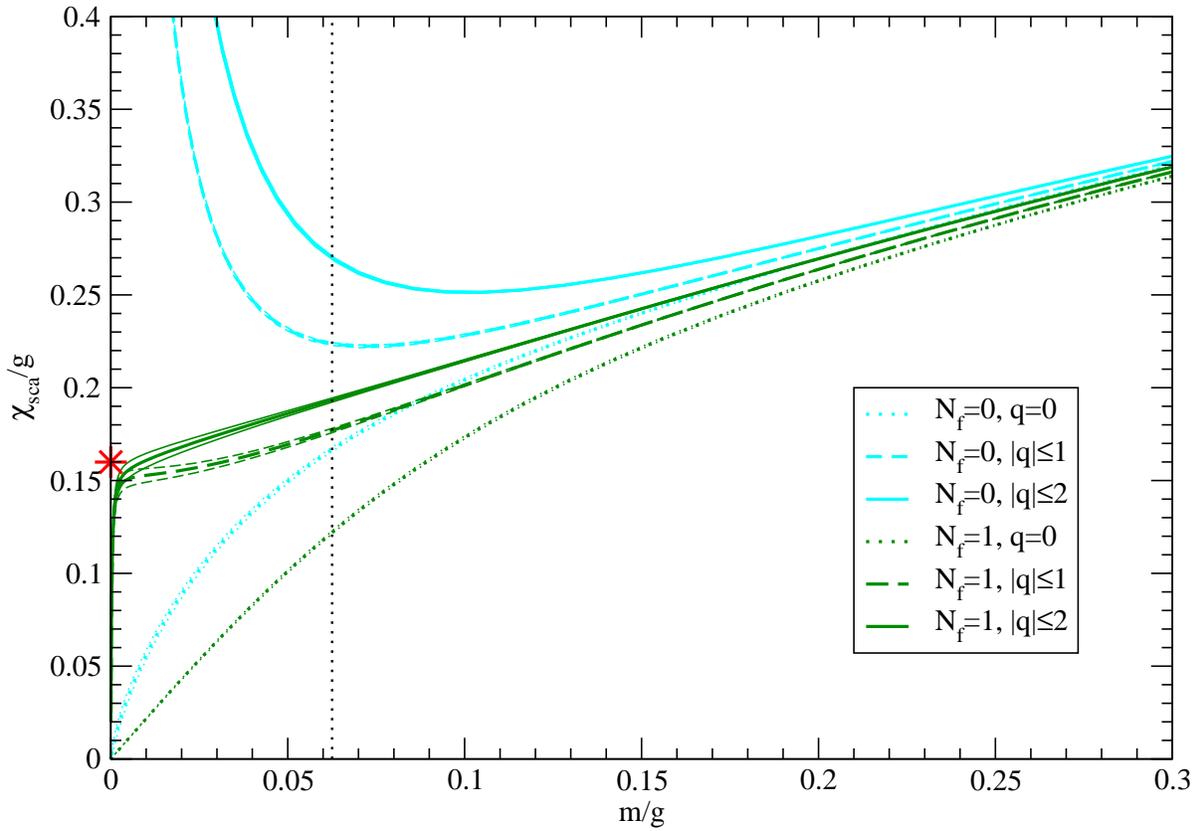}
\end{center}
\vspace{-6mm}
\caption{Sectoral condensate for staggered quarks with 1 and 3 steps of APE/HYP
smearing.}
\label{fig:secs}
\end{figure}

Fig.\,\ref{fig:seco} displays our results for the condensate if we truncate the
partition function at a given $|q|_\mr{max}$.
The overlap construct faithfully reproduces the ``selection theorem'', which
means that the $\Nf\!=\!1$ condensate tends to zero if one restricts it to the
topologically trivial sector, while it takes the Schwinger value
(\ref{schwinger}) in the chiral limit for any other $|q|_\mr{max}$.

Analogously, the staggered condensate exhibits (for both $\Nf$ shown) a
many-sigma difference between $|q|_\mr{max}\!=\!0,1,2$.
This disproves the widely believed fiction that staggered fermions are
``insensitive to topology'' -- but they are \emph{not sensitive the right way};
they don't seem to know about the index and the vanishing theorems which are at
the root of the ``selection theorem''.

\bigskip

Fig.\,\ref{fig:secs} shows that the real surprise comes again after one or a
few APE/HYP-smearing steps.
Already one step lets the staggered $\Nf\!=\!1$ condensate (at intermediate
mass) develop a marked sensitivity on the topological charge of the background.

After two more steps the qualitative picture is just like in the (unsmeared)
overlap case (c.f.\ Fig.\,\ref{fig:seco}), i.e.\ \emph{smeared staggered
fermions do know about the relationship between topology and the chirality of
zero-modes} -- down to rather small (but non-zero) quark masses.

Of course, if one really performs the chiral limit, the staggered condensate
still tends to zero -- for any $|q|_\mr{max}$ and even in the quenched case
($\Nf\!=\!0)$.
This is visible in the top of Fig.\,\ref{fig:secs}, while in the bottom part
it is off the scale.

\bigskip

Since in the Schwinger model various observables have been seen to depend on
the topological charge \cite{previoustopology}, it is surprising that the
``selection theorem'' has not been checked before.


\section{Topological susceptibility}

Another interesting observable to study the effects of dynamical fermions is
the topological susceptibility which, in the context of this note, shall be
\emph{defined} (in the continuum) through
\beq
\ch_\mr{top}=\lim_{V\to\infty}
{\<\det(D\!+\!m)^\Nf\;q^2\>\ovr V\,\<\det(D\!+\!m)^\Nf\>}
\;.
\label{defsusc}
\eeq
The main difference to the scalar condensate is that the topological
susceptibility depends only on the \emph{sea}-quarks, thus offering a
potentially cleaner view at the effects of square-rooting the staggered
determinant to get $\Nf\!=\!1$.

For staggered quarks, the definition (\ref{defsusc}), taken in fixed volume,
reduces to
\beq
{\ch_\mr{top}\ovr g^2}=
{\be\ovr N^2}\,{\<\det(D^\mr{st}_m)^{\Nf/2}\;q^2\>\ovr
\<\det(D^\mr{st}_m)^{\Nf/2}\>}
\;,
\label{suscstag}
\eeq
while for overlap quarks, the implementation reads
\beq
{\ch_\mr{top}\ovr g^2}=
{\be\ovr N^2}\,{\<\det(D^\mr{ov}_m)^\Nf\;q^2\>\ovr
\<\det(D^\mr{ov}_m)^\Nf\>}
\;,
\label{suscover}
\eeq
where $\det(D^\mr{st}_m)$ and $\det(D^\mr{ov}_m)$ are defined in
(\ref{condstageig}, \ref{condovereig}).
In either case the sum is over the spectrum of the massless operator, and we
apply the overlap definition (\ref{defind}) of the topological charge both in
(\ref{suscstag}) and in (\ref{suscover}).

Finally we like to mention the continuum prediction how the topological
susceptibility in QCD (!) tends to zero, if the quark mass does
\cite{Leutwyler:1992yt} (the LS parameter $x$ was defined in (\ref{lsp})),
\bea
\ch_\mr{top}&=&{\Sigma m\ovr\Nf} \qquad (\Nf\!=\!1 \vee x\!\gg\!1)
\label{suscinfvol}
\\
\ch_\mr{top}&\propto&m^\Nf \qquad\qquad\qquad\! (x\!\ll\!1)
\label{suscfinvol}
\;.
\eea

\bigskip

\begin{figure}
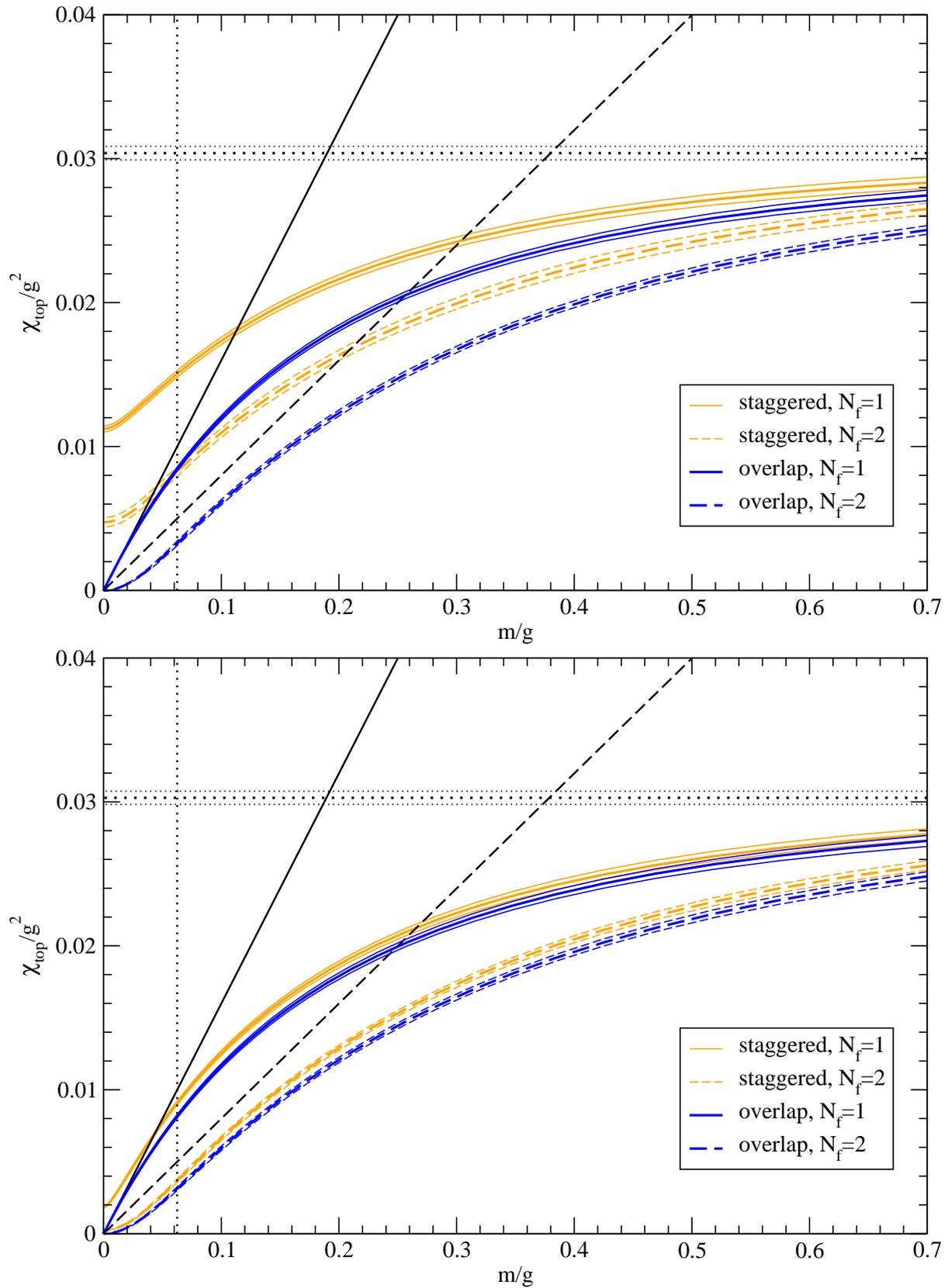

\begin{center}
\epsfig{file=smrw_anasusc0.eps,height=11cm}
\epsfig{file=smrw_anasusc1.eps,height=11cm}
\end{center}
\vspace{-6mm}
\caption{Topological susceptibility with 0 (top) and 1 steps of APE/HYP
smearing.}
\label{fig:susc}
\end{figure}

Fig.\,\ref{fig:susc} contains our results for the mass dependence of the
topological susceptibility, defined via (\ref{suscstag}, \ref{suscover}).
The full and the dashed lines represent the prediction of (\ref{suscinfvol})
for the cases $\Nf\!=\!1,2$, respectively.
One can see that for both discretizations $\ch_\mr{top}$ gets suppressed by
dynamical fermion effects, but close to the chiral limit only the overlap
determinant leads to results which are compatible with the QCD prediction
(\ref{suscinfvol}) for $\Nf\!=\!1$ and (\ref{suscfinvol}) for $\Nf\!=\!2$.

Again, just one smearing step proves sufficient to almost eliminate the lattice
artefacts the ``unsmeared/naive'' staggered determinant was plagued with, while
the topological susceptibility for overlap fermions stays basically invariant
under such a modification.


\section{Spectral hint}

\begin{figure}
\begin{center}
\epsfig{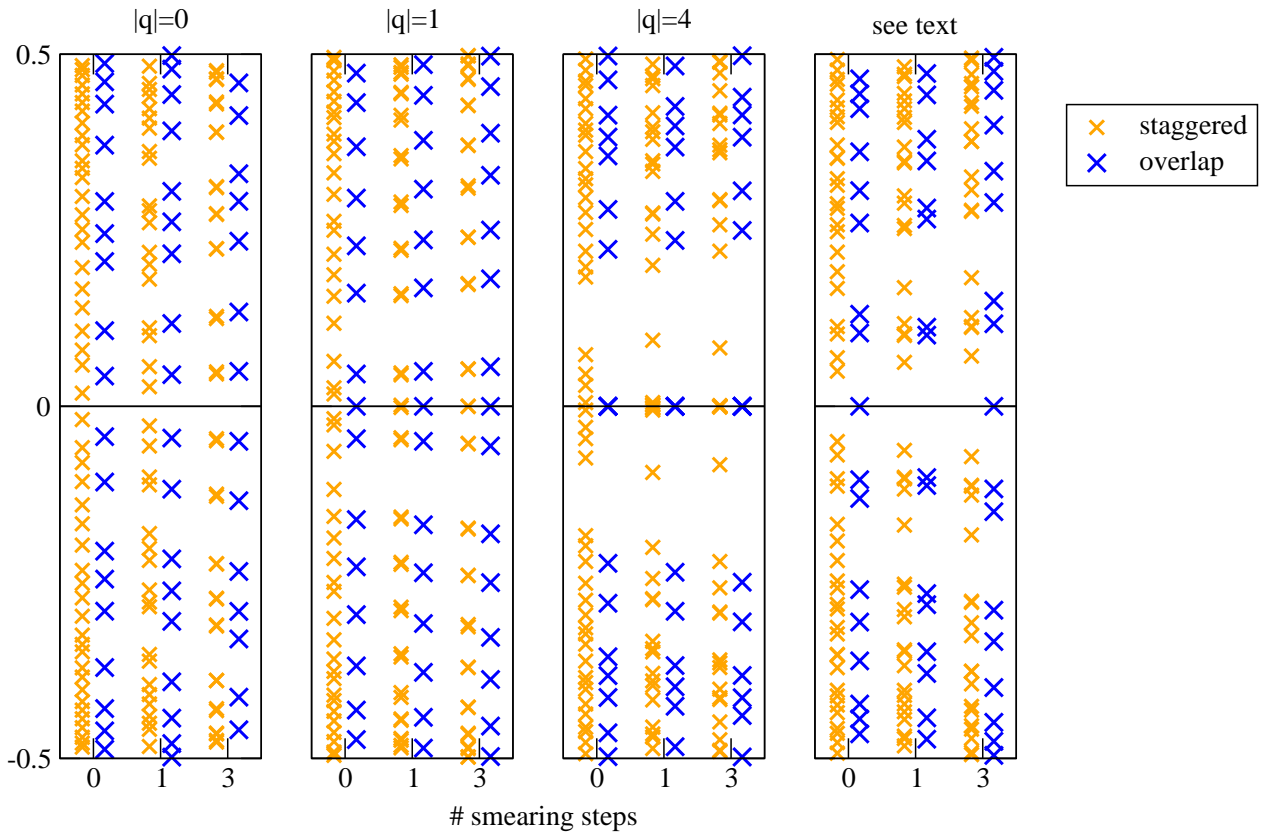}
\end{center}
\vspace{-6mm}
\caption{Physically relevant region of the Dirac spectrum on some selected
configurations.}
\label{fig:spechint}
\end{figure}

As a hint of what is the likely reason behind the remarkable success of one or
a few APE/HYP smearing steps at the $\be$-value considered, we like to present
the effect of such a modification on the spectrum of some individual
configurations.
In Fig.\,\ref{fig:spechint} we plot the physically relevant part of the
spectrum of the staggered (the $a\la$ from (\ref{condstageig})) and overlap
(the $a\hat\la$ from (\ref{condovereig2})) operator on four selected
configurations before and after smearing.

The two configurations on the left are typical examples for topological charge
$q=0$ and $|q|=1$, respectively.
One sees that on the unsmeared configurations the staggered spectrum does not
resemble that of the overlap operator.
Furthermore, it is hard to see a qualitative difference between $q\!=\!0$ and
$|q|\!=\!1$ in the spectrum of the original staggered operator.
After a few smearing steps this picture changes.
The eigenvalues of the staggered operator seem to form near-degenerate pairs
which sit close to a single overlap $a\hat\la$ on the same configuration.
In particular, in the $|q|\!=\!1$ case a pair of eigenmodes moves very close
to the real axis.
Clearly, such a shift mimics the effect of the true zero-mode in the overlap
counterpart down to rather small quark masses.
It is only when the mass becomes comparable to the smallest (smeared staggered)
eigenvalue that the absence of an exact zero-mode matters and a qualitative
difference between the staggered and overlap Dirac operators (on that
configuration) shows up.

In the third column of Fig.\,\ref{fig:spechint}, a typical configuration with
higher topological charge ($|q|\!=\!4$) is displayed.
Here, the picture is not so nice any more, since after $3$ smearing steps only
$3$ pairs of eigenmodes have come close to the real axis and the fourth one is
still somewhat further out.

To show that these findings are not generic, the last column of
Fig.\,\ref{fig:spechint} presents a selected configuration on which the
topological charge varies repeatedly under subsequent smearing steps.
Qualitative resemblance between the spectra of the two operators is vague at
best and there is no evidence of a pair of staggered eigenvalues moving close
to the real axis.
Obviously, such ``sick'' configurations will occur more frequently at larger
coupling.


\section{Summary}

Our findings may be summarized by the following statements:
\begin{enumerate}
\item
At finite spacing and in a finite box, the ``naive'' staggered action leads to
a scalar condensate which is \emph{qualitatively wrong}: the staggered results
vanish in the chiral limit for any $\Nf$, while the overlap successfully
reproduces the quenched $\sim\!1/m$ divergence, and the analytically known
Schwinger value in the chiral limit for $\Nf\!=\!1$.
\item
Considering both types of ``hybrid'' formulations (staggered valence quarks
with overlap sea quarks and vice versa) we find that the failure of the naive
staggered formulation cannot be attributed to either the determinant or the
propagator alone.
\item
The ``selection theorem'' is reproduced, in an impressive manner, with
overlap fermions, while naive staggered fermions fail completely.
\item
The topological susceptibility shows lattice artefacts which are large for
unimproved staggered fermions, while the overlap results seem consistent with
the known chiral behavior.
\item
Taking the square root of the staggered determinant to have $\Nf\!=\!1$ seems
to be no more or less harmful than dividing by 2, in the valence sector, to
get the 1-flavor condensate.
All naive staggered results seem to only gradually vary for $\Nf\!=\!0,1,2$ --
even near the chiral limit, where they shouldn't.
\item
At the $\be$ value considered, already one smearing step brings a remarkable
improvement: although formally the chiral limit of staggered fermions is still
wrong, the mass at which one begins to see this is dramatically lowered.
At moderate quark masses, one observes a clear blow-up of the APE/HYP smeared
staggered condensate for $\Nf\!=\!0$, while it stays close to the Schwinger
value (\ref{schwinger}) for $\Nf\!=\!1$ and keeps the qualitatively correct
behavior at $\Nf\!=\!2$.
Moreover, in spite of naive staggered fermions being misguided in the way they
see topology, the APE/HYP variety condensate shows a remarkable sensitivity on
$|q|_\mr{max}$ for $\Nf\!=\!0$ and $\Nf\!=\!1$, reproducing the ``selection
theorem'' in the latter case.
Finally, the staggered artefacts in the topological susceptibility get
drastically reduced, resulting in good agreement with the overlap curve down
to very small quark masses.
\item
The main difference after a few smearing steps is the size of lattice artefacts
in the condensate at large quark masses which make the smeared overlap
condensate (for any $\Nf$) lie closer to the analytical prediction
(\ref{smilga}) by Smilga, while the smeared staggered values move more towards
the free field prediction (\ref{condstagfree}).
The qualitative failure of staggered fermions in the chiral limit is only
visible at very small quark masses.
\item
Under moderate smearing the modes of the massless staggered Dirac operator form
near-degenerate pairs which move close to the corresponding overlap (single)
eigenvalues ($\hat\la$), supporting the square root and factor $1/2$ procedure
for staggered spectroscopy.
On non-trivial backgrounds the mismatch between the staggered fake zero-mode
and the true overlap zero gives an estimate of the quark mass at which the
deficiency of the staggered formulation gets visible, though there are
configurations which break this analogy.
\end{enumerate}

\noindent
As a warning against an overly optimistic interpretation of 6/8, we feel
obliged to recall that we were working with fairly smooth gauge fields.
It is not clear whether such a nice pattern holds true also in QCD at
accessible couplings.

\bigskip

{\bf Acknowledgments:}
We are indebted to Gunnar Bali for bringing to our attention the interest in
checking chiral properties of staggered fermions in a simple model. We thank
Steve Sharpe for the suggestion to study the effect of APE/HYP smearing.
S.D.\ wishes to acknowledge useful discussions with Rainer Sommer, C.H.\ with
Christian Lang and Laurent Lellouch. S.D.\ is supported by DFG in SFB/TR-9,
C.H.\ is supported by EU grant HPMF-CT-2001-01468.


\end{document}